\documentclass[preprint,aps,prd,showpacs,nofootinbib,superscriptaddress,floatfix]{revtex4-1}

\usepackage{amsmath}
\usepackage{graphicx}
\usepackage{epstopdf}
\usepackage{epsfig}
\usepackage{amssymb}
\usepackage{bm}
\newcommand{\beq}{\begin{equation}}
\newcommand{\eeq}{\end{equation}}

\usepackage{color}

\allowdisplaybreaks[1]

\begin{document}

\begin{flushright}
\end{flushright}

\title{Spontaneous SUSY breaking in natural $SO(10)$ grand unified theory}

\author{Nobuhiro Maekawa}
\email[]{maekawa@eken.phys.nagoya-u.ac.jp}
\affiliation{Department of Physics,
Nagoya University, Nagoya 464-8602, Japan}
\affiliation{Kobayashi-Maskawa Institute for the Origin of Particles and the
Universe, Nagoya University, Nagoya 464-8602, Japan}

\author{Yuji Omura}
\email[]{yomura@phys.kindai.ac.jp}
\affiliation{Department of Physics, Kindai University, Higashi-Osaka, Osaka 577-8502, Japan}

\author{Yoshihiro Shigekami}
\email[]{sigekami@post.kek.jp}
\affiliation{School of Physics, Huazhong University of Science and Technology, 
Wuhan 430074, China}

\author{Manabu Yoshida}
\email[]{manabu@eken.phys.nagoya-u.ac.jp}
\affiliation{Department of Physics,
Nagoya University, Nagoya 464-8602, Japan}

\date{\today}


\begin{abstract}
\noindent
We propose a simple grand unified theory (GUT) scenario in which supersymmetry (SUSY) is 
spontaneously broken in visible sector. 
Our model is based on the GUT model that has been proposed to solve almost all problems in conventional GUT scenarios. In the previous work, the problems can be solved by a natural assumption in a supersymmetric vacuum. In this paper, we consider an extension of the model (i.e. omitting one singlet field) and break SUSY spontaneously without new sector. 
Our model does not have hidden sector and predicts
high-scale SUSY where sfermion masses are of order 100-1000 TeV and flavor violating processes are suppressed. In this scenario, we can see an explicit signature of GUT in sfermion mass spectrum since the sfermion mass spectrum respects $SU(5)$ matter 
unification. 
In addition, we find a superheavy long lived charged lepton as a proof of our scenario, and it may be seen in the LHC.
\end{abstract}


\maketitle

\section{Introduction}
\label{sec:intro}
Grand unified theory (GUT) \cite{GG} realizes two kinds of unification. Three gauge groups 
$SU(3)_C\times SU(2)_L\times U(1)_Y$ in the standard model (SM) can be unified into a single
group $SU(5)$, $SO(10)$, or $E_6$, which leads to the unification of forces. In addition, one generation quarks and leptons are unified into two multiplets
$\bf 10$ and $\bf\bar 5$ in $SU(5)$ GUT. In $SO(10)$ GUT, moreover, these fields as well as the right-handed 
neutrino are unified into a single multiplet $\bf 16$, which results in the unification of matters. 
For both unifications, we have experimental evidences. For the unification
of forces, three gauge couplings in the SM meet at a scale when supersymmetry
(SUSY) is introduced around the weak scale. 
For the unification of matters, observed hierarchies of quark and lepton masses and
mixings can be understood by a simple assumption that unified matter fields $\bf 10$ of $SU(5)$ induce
stronger hierarchies for Yukawa couplings than $\bf\bar 5$ fields. 
Moreover, SUSY plays an important role in the GUT scenario. 
SUSY avoids the fine-tuning in Higgs masses, and the lightest SUSY particle can be the dark matter (DM) in addition
to the success of the gauge coupling unification.
Therefore, SUSY GUT is the most
promising candidate as a model beyond the SM. 

Unfortunately, SUSY GUT is suffering from several problems. 
The most serious one is called the doublet-triplet splitting problem \cite{DTS}. A fine-tuning is needed to obtain the weak 
scale mass of the SM Higgs (doublet Higgs) and the colored Higgs (triplet Higgs) mass larger than the GUT scale because the doublet Higgs and triplet Higgs are included in the same multiplet $\bf 5$ of $SU(5)$. 
Another problem is that the matter unification results in unrealistic Yukawa relations in simple GUT models.
In the minimal $SU(5)$ GUT, the down quark Yukawa matrix $Y_d$ is equal to the transposed matrix of
the charged lepton Yukawa matrix $Y_e$ as $Y_d=Y_e^t$. In the minimal $SO(10)$ GUT, all the Yukawa matrices
become equal as $Y_u=Y_d=Y_e=Y_{\nu_D}$, where $Y_u$ and $Y_{\nu_D}$ are Yukawa matrices of the up quark and
the Dirac neutrino. These Yukawa relations are inconsistent with the observed quark and lepton masses, so
some improvements are needed to obtain realistic Yukawa couplings.

There are several problems related with SUSY, which are in principle independent of the above GUT
problems. SUSY must be broken because we have no SUSY at
low energy.
In addition, generic SUSY breaking parameters induce too large flavor changing neutral current (FCNC) processes and too large
CP violating processes if SUSY breaking scale is just above the weak scale to stabilize the weak scale.
Thus, we have to find out the way to control a lot of  SUSY breaking parameters in the
minimal SUSY SM (MSSM).
In many scenarios, we introduce hidden sector, in which SUSY is spontaneously broken, and mediation sector, where there are some fields to
mediate SUSY breaking effects to visible sector.
 One reason to introduce
these complicated sectors is to realize universal sfermion masses via universal interaction in mediation sector, and these can suppress the FCNC processes.

In this paper, we propose a scenario in which SUSY is spontaneously broken in the visible sector. 
We have neither hidden sector nor mediation sector. We adopt namely
the natural GUT \cite{naturalGUT} as the visible sector. In the natural GUT, all the above problems on GUT can be solved under the natural
assumption that all (higher dimensional) interactions are taken into account with O(1) coefficients. The anomalous $U(1)_A$ gauge symmetry \cite{anomalousU(1)} plays an important role in the natural GUT. On the other hand, it is known that
the Fayet-Illiopoulos (FI) term \cite{FI}, which is induced by the anomalous $U(1)_A$ gauge symmetry, can play a critical role in breaking SUSY spontaneously. Therefore, we study the possibility that SUSY is spontaneously broken
in the natural GUT. Interestingly, the scenario predicts high-scale SUSY which suppress the FCNC processes without the hidden sector.


In Sec.~\ref{sec:nGUTreview}, we review the natural $SO(10)$ GUT which gives a new explanation for the success of gauge coupling unification in minimal $SU(5)$ GUT. 
In Sec.~\ref{sec:SSB}, we propose a natural $SO(10)$ GUT in which SUSY is spontaneously broken. We also discuss the phenomenology of our scenario in this section.
Sec.~\ref{sec:summary} is devoted to summary and discussion.

\section{Brief review of the natural GUT}
\label{sec:nGUTreview}
In the natural GUT in which anomalous $U(1)$ gauge symmetry plays an important role, 
various problems, e.g. the doublet-triplet splitting problem, can be
solved under the natural assumption. In this model, all interactions including higher 
dimensional interactions are introduced with $O(1)$ coefficients. 
Once we fix the symmetry in the model, we can define the theory except $O(1)$ coefficients. 
Under this natural assumption, we have to control infinite number of higher dimensional
interactions. This is possible by using the SUSY (holomorphic) zero mechanism.
In this section, we briefly review the natural GUT. We explain how to control the infinite 
number of interactions and to solve the doublet-triplet splitting problem.

\subsection{How to determine the vacuum expectation values}
\label{sec;breaking}

The anomalous $U(1)_A$ gauge symmetry is a $U(1)$ gauge symmetry with gauge anomaly
which is cancelled by the Green-Schwarz mechanism \cite{GS}. Nevertheless in this paper, we use anomalous
$U(1)_A$ theory as just a $U(1)$ gauge theory with FI term,
$\xi^2\int d\theta^2V_A$, where $V_A$ is a vector multiplet of the $U(1)_A$.
One of the most important features of the models with anomalous $U(1)_A$ gauge symmetry is that the vacuum expectation values (VEVs) are determined by their
$U(1)_A$ charges as
\begin{equation}
\langle Z\rangle \sim \left\{\begin{array}{ll} 0 & (z>0) \\
                                                         \lambda^{-z} & (z\leq 0)
                                                         \end{array}\right.,
\label{VEV}
\end{equation}
where $\lambda\equiv \xi/\Lambda\ll 1$, $z$ is the $U(1)_A$ charge of the field $Z$, and
$\Lambda$ is the cutoff scale which is usually taken to be 1 in this paper. In the followings,
we use large characters for fields or operators and small characters for their $U(1)_A$ charges. These VEVs determine the coefficients of interactions, for example,
Yukawa interaction $XYZ$, except the $O(1)$ coefficients as
\begin{equation}
\left\{\begin{array}{cl}
 \lambda^{x+y+z}XYZ & (x+y+z\geq 0) \\
  0 & (x+y+z<0)
\end{array}\right.
\end{equation}
if we assume that all interactions which are allowed by the $U(1)_A$ gauge symmetry have 
$O(1)$ coefficients. Note that interactions which have negative $U(1)_A$ charges are
forbidden, which is called the SUSY zero mechanism (or holomorphic zero mechanism).
This feature plays an important role in solving the doublet-triplet splitting problem.
We explain the above features by several examples. 

If we have only one negatively charged field $\Theta$ whose 
$U(1)_A$ charge is $\theta=-1$, then the VEV is fixed by $D$ flatness condition of
$U(1)_A$ gauge symmetry,
\begin{equation}
D_A=\frac{g_A}{2}(\xi^2-|\Theta|^2)=0,
\end{equation}
as $\langle \Theta\rangle=\lambda\Lambda$. Then the interaction $XYZ$ is obtained from
the $U(1)_A$ invariant interaction as
\begin{equation}
\left(\frac{\Theta}{\Lambda}\right)^{x+y+z}XYZ\rightarrow \lambda^{x+y+z}XYZ
\end{equation}
by developing the VEV, $\langle \Theta\rangle$. Obviously, if $x+y+z<0$, then such 
interaction is forbidden. Note that if a field $A$ has non vanishing VEV as
$\langle A\rangle\sim \lambda^{-a}$, then the higher dimensional interaction 
$\lambda^{x+y+z+a}XYZA$ gives the same order of the coefficients for the interaction
$XYZ$ as 
\begin{equation}
\lambda^{x+y+z+a}XYZ\langle A\rangle\sim\lambda^{x+y+z}XYZ.
\end{equation}
Therefore, higher dimensional interactions give the similar contributions to the coefficients
of the interactions, which is important in avoiding unrealistic GUT relations between
Yukawa couplings when $A$ is the adjoint Higgs.

Let us consider only singlets $Z_i^+$ and $Z_j^-$ ($i=1,2,\cdots, n_+$, $j=1,2,\cdots, n_-$) under GUT group for simplicity. Here, the $U(1)_A$
charges for them are $z_i^+>0$ and $z_j^-<0$. Generically, $n_++n_--1$  $F$ flatness
 conditions in complex number and one $D$ flatness condition in real number determine
 $n_++n_-$ complex VEVs except one real freedom by the $U(1)_A$ gauge symmetry.
 Here $-1$ in the number of $F$ flatness condition is caused by the gauge invariance of the
 superpotential. If all interactions which are allowed by $U(1)_A$ gauge symmetry are
 introduced with $O(1)$ coefficients, VEVs of these singlets must be $O(1)$ generically.
 As another possibility, let us consider that all positively charged fields $Z_i^+$ have
 vanishing VEVs. Then, the $F$ flatness conditions of negatively charged fields are 
 automatically satisfied, and therefore, $n_+$ $F$ flatness conditions of positively
 charged fields and a $D$ flatness condition determine the VEVs of negatively charged
 fields. The situation is generically given by
 \begin{equation}
 \left\{\begin{array}{ll}
 n_+>n_--1 :& {\rm Overdetermined.\ SUSY\ is\ spontaneously\ broken\ in\ meta\mathchar`-stable\ vacua} \\
 n_+=n_--1 :& {\rm All\ VEVs\ are\ fixed\ and\ no\ flat\ direction} \\
 n_+<n_--1 :& {\rm Flat\ directions\ and\ massless\ modes\ appear}
 \end{array}\right.
 \end{equation}
 To fix the VEVs of negatively charged fields, it is sufficient to consider 
 the superpotential which includes only one positively charged field.
 Therefore, the number of important terms for fixing VEVs become finite
 because of the SUSY zero mechanism
 although infinite number of higher dimensional terms are introduced.   
 
\subsection{Higgs sector in the natural $SO(10)$ GUT}
The minimal Higgs content that breaks $SO(10)$ into the standard gauge group is
one adjoint Higgs $A({\bf 45})$ and one pair of spinor Higgs 
$C({\bf 16})+\bar C({\bf\overline{16}})$. The standard model Higgs is included in $H({\bf 10})$. They must
have negative $U(1)_A$ charges because they have non-vanishing VEVs. To fix the VEVs,
we have to introduce the same number of positively charged fields
$A'({\bf 45})$, $C'({\bf 16})+\bar C'({\bf\overline{16}})$, and $H'({\bf 10})$. 
This is a minimum content for $SO(10)$ GUT with anomalous $U(1)_A$ gauge symmetry.
Interestingly, this minimal Higgs contents (+several singlets) can solve the doublet-triplet
splitting problem which is the most serious problem in SUSY GUT scenario.
Quantum numbers of these Higgs fields and matter fields (three ${\bf 16}$ and one ${\bf 10}$)
are shown in Table \ref{SO10Higgs}. Note that the half integer $U(1)_A$ charges for matter fields
play the same role as R-parity. Since the half integer $U(1)_A$ charges are positive, the $F$ flatness
conditions of matter fields are automatically satisfied.  
In this $SO(10)$ model, the doublet-triplet splitting in addition to realistic quark and lepton masses and mixings can be realized under the natural assumption explained above. 
In other words, we have fixed the $U(1)_A$ charges so that these constraints are satisfied.
(For the details, see refs. \cite{naturalGUT,gcu}.)
Note that although $n_+=n_--2$ in Higgs sector, no flat direction and massless mode appear
except a pair of Higgs doublets. This looks to be inconsistent with the above general 
arguments. This is because some modes including one SM singlet are
absorbed by the Higgs mechanism.
\begin{table}[t]
  \begin{center}
    \begin{tabular}{|c||c|c|c|}
      \hline
      $SO(10)$ & negatively charged fields & positively charged fields & matter fields \\ \hline \hline
      {\bf 45}  & $A(a=-1,-)$ & $A'(a'=3,-)$ & \\ \hline
      {\bf 16} & $C(c=-4,+)$ & $C'(c'=3,-)$  & $\Psi_i(\psi_1=\frac{9}{2}, \psi_2=\frac{7}{2}, \psi_3=\frac{3}{2}, +)$ \\ \hline
      ${\bf \overline{16}}$ & $\bar C(\bar c=-1,+)$ & $\bar C'(\bar c'=7,-)$ & \\ \hline
      {\bf 10} & $H(h=-3,+)$  & $H'(h'=4,-)$ & $T(t=\frac{5}{2},+)$ \\ \hline
      1 &  $\left.\begin{array}{c}\Theta(\theta=-1,+), \\ Z(z=-2,-), \bar Z(\bar z=-2,-) \end{array}\right.$ & $Z'(z'=5,+)$ &
      \\ \hline
    \end{tabular}
    \caption{Field contents of natural $SO(10)$ GUT with $U(1)_A$ charges. $\pm$ shows $Z_2$ parity. 
     The half integer $U(1)_A$ charges play the same role as R-parity.}
     \label{SO10Higgs}
  \end{center}
\end{table}

To fix the VEVs of the negatively charged fields, it is sufficient to consider the superpotential $W_{X'}$, which is linear in one positively charged field 
$X'=A',C',\bar C',H',Z'$.  Let us examine $W_{X'}$ one by one.

First, we consider $W_{A'}=\lambda^{a'+a}{\rm tr}(A'A)+\lambda^{a'+3a}({\rm tr}(A'A^3)
+{\rm tr}(A'A){\rm tr}(A^2))$. In this paper, we omit the $O(1)$ coefficients in the 
superpotential. Without loss of generality, the VEV of the adjoint field can be written
as 
\begin{equation}
\langle A\rangle=\left(\begin{array}{cc} 0 & 1 \\ -1 & 0 \end{array}\right)\otimes\left(\begin{array}{ccccc}x_1 & 0 & 0 & 0 & 0 \\
                                                                  0 & x_2 & 0 & 0 & 0 \\
                                                                  0 & 0 & x_3 & 0 & 0 \\
                                                                  0 & 0 & 0 & x_4 & 0 \\
                                                                  0 & 0 & 0 & 0 & x_5\end{array}\right).
\end{equation}
The $F$ flatness condition $\partial W_{A'}/\partial A'=0$ leads to
$x_i(\lambda^{a'+a}+\lambda^{a'+3a}x_i^2)=0$, which gives two solutions $x_i=0,V$.
Here $V\sim \lambda^{-a}$. The vacua can be classified by the number of 0 components.
If it is 2, we can obtain the Dimopoulos-Wilczek (DW) type VEV \cite{DW}, which is important in
solving the doublet-triplet splitting problem. Note that terms $\lambda^{a'+na}A'A^n (n>3)$
are forbidden by the SUSY zero mechanism. If they are allowed, it becomes less natural
to obtain the DW type VEV. This VEV breaks $SO(10)$ into $SU(3)_C\times SU(2)_L\times SU(2)_R\times U(1)_{B-L}$. 

Second, we consider $W_{H'}=\lambda^{h'+h+a}H'AH$. The term $H'H$ is forbidden by
$Z_2$ parity. 
This term gives the mass $\lambda^{h'+h}$
to the triplet Higgs but not the doublet Higgs under the DW VEV of $A$. If we include
the mass term $\lambda^{2h'}H'^2$, only one pair of doublet Higgs is massless, and 
therefore, the doublet-triplet splitting can be realized. The effective triplet Higgs mass
for nucleon decay can be estimated as $m_{eff}\sim \lambda^{2h}$ which is larger than
the cutoff scale because of negative $h$, and hence, the dimension 5 proton decay is sufficiently suppressed. 

Third, we consider $W_{Z'}=\lambda^{z'}Z'(1+\lambda^{\bar c+c}\bar CC)$, where the first
term includes the contributions from $Z$, $\bar Z$, and $A$. The $F$ flatness condition
of $Z'$ leads to $\langle \bar CC\rangle\sim\lambda^{-(\bar c+c)}$, and therefore, we
obtain $\langle \bar C\rangle=\langle C\rangle\sim\lambda^{-\frac{1}{2}(\bar c+c)}$ due
to the $D$ flatness condition of $SO(10)$. This VEV is expected to break $SU(2)_R\times U(1)_{B-L}$ into $U(1)_Y$, but it is not guaranteed generically. Fortunately, an alignment
mechanism is embedded in this model as discussed in the next paragraph.

Finally, we consider $W_{\bar C'}=\lambda^{\bar c'+c}\bar C'(\lambda^zZ+\lambda^aA)C$
and $W_{C'}=\lambda^{\bar c+c'}\bar C(\lambda^{\bar z}\bar Z+\lambda^aA)C'$, which can realize
the alignment. This mechanism  was proposed by Barr and Raby \cite{Barr:1997hq}, and we call it the Barr-Raby mechanism in this paper. 
We examine $W_{\bar C'}$ here. ($W_{C'}$ gives the similar result.)
An important point is that the VEV of the adjoint Higgs is proportional to the $B-L$ charge. 
Therefore, at least one of the components of $C$ must have non-vanishing VEV as
$\langle C_f\rangle\neq 0$, where $f$ is one of the components of $C$ which is 
divided into 
$({\bf 3}, {\bf 2}, 1)_\frac{1}{3}+({\bf\bar 3}, 1, {\bf 2})_{-\frac{1}{3}}+(1,{\bf 2},1)_{-1}+(1,1,{\bf 2})_1$ under $SU(3)_C\times SU(2)_L\times SU(2)_R\times U(1)_{B-L}$.
Then $F$ flatness condition of $\bar C'$ leads to $(\lambda^z Z+q_f)C_f=0$ 
which fixes the VEV of $Z$. Here, $q_f$ is $B-L$ charge of the component $C_f$.
Then the other component fields $C_{f'} (f'\neq f)$ have vanishing VEVs because of
$F$ flatness conditions $(\lambda^zZ+q_{f'})C_{f'}=0$ and  $(\lambda^zZ+q_{f'})\neq 0$
due to $q_{f'}\neq q_f$. Therefore, alignment is realized. If $f=(1,1,{\bf 2})_1$\footnote{Even in the case $f=(1,{\bf 2},1)_{-1}$, the SM gauge group can be obtained by exchanging the names of $SU(2)_L$ and $SU(2)_R$.}, we can obtain the SM gauge group. 

Here, we do not show the mass spectrum of this Higgs sector explicitly, but all fields 
except one pair of Higgs doublet become superheavy in this model. 

\subsection{Matter sector in the natural SO(10) GUT}
Basically, Yukawa couplings can be obtained from the interactions
\begin{equation}
W=\sum_{i,j}^3\lambda^{\psi_i+\psi_j+h}\Psi_i\Psi_jH
+\sum_i^3\lambda^{\psi_i+t+c}\Psi_iTC+\lambda^{2t}T^2,
\end{equation}
where $O(1)$ coefficients are neglected and higher dimensional interactions
like $\lambda^{\psi_i+\psi_j+2a+h}\Psi_iA^2\Psi_jH$ avoid unrealistic GUT 
relations in Yukawa couplings after developing the VEVs, for example, 
$\langle A\rangle\sim \lambda^{-a}$. It is important that three massless 
${\bf\bar 5}$ fields which includes the SM quarks and leptons become 
\begin{equation}
({\bf\bar 5}_1, {\bf\bar 5}_2, {\bf\bar 5}_3)\sim ({\bf\bar 5}_{\Psi_1}, {\bf\bar 5}_T, {\bf\bar 5}_{\Psi_2})
\end{equation}
 since ${\bf\bar 5}_{\Psi_3}$ in $\Psi_3$ becomes superheavy with 
${\bf 5}_T$ in $T$. This structure is important in obtaining realistic quark 
and lepton mass matrices.

\subsection{Gauge coupling unification \cite{gcu}}
The mass spectrum and VEVs are fixed by their anomalous $U(1)_A$ charges, but unfortunately
the mass spectrum does not respect $SU(5)$ symmetry. For example, the masses of
the adjoint Higgs of $SU(5)$ are basically determined by the mass term 
$\lambda^{a'+a}A'A$ as $\lambda^{a'+a}$. However, since a component field $({\bf 3}, {\bf 2})_{-\frac{5}{6}}$ in $A$ is absorbed by the Higgs mechanism, the corresponding field
in $A'$ has no partner in $A$, and therefore, its mass becomes $\lambda^{2a'}$ which
is obtained from the mass term $\lambda^{2a'}A'^2$. Obviously, this mass spectrum does
not respect $SU(5)$, and therefore, it may spoil the success of gauge coupling unification
in the minimal SUSY $SU(5)$ GUT. 

Interestingly, although the mass spectrum of superheavy Higgs sector does not respect $SU(5)$, 
the natural GUT can explain the success of gauge coupling unification in the minimal SUSY $SU(5)$ GUT.
Since in the natural GUT, all the mass scales and VEVs except $O(1)$ coefficients are determined by anomalous $U(1)_A$ charges, 
we can calculate the renormalization group equations (RGEs) when all the $O(1)$ coefficients are fixed, for
example, as one.
When three gauge couplings $\alpha_i\equiv g_i^2/4\pi$ $(i=1,2,3)$ at the SUSY breaking scale 
$\Lambda_{SUSY}$ are given by
\begin{equation}
\alpha_i^{-1}(\Lambda_{SUSY})=\alpha_G^{-1}(\Lambda_G)+\frac{1}{2\pi}\left(b_i\ln\left(\frac{\Lambda_G}{\Lambda_{SUSY}}\right)\right),
\end{equation}
the two conditions for gauge coupling unification $\alpha_1(\Lambda)=\alpha_2(\Lambda)=\alpha_3(\Lambda)$
can be rewritten by two relations $\Lambda=\Lambda_G$ and $h=0$. Here, $\alpha_G$, $\Lambda_G$ and $b_i$ are the unified gauge coupling in the minimal $SU(5)$ GUT, 
the usual unification scale and the renormalization group coefficients in the MSSM, respectively. Surprisingly, all anomalous $U(1)_A$ charges except the SM Higgs' charge are cancelled out in the conditions. The first condition $\Lambda=\Lambda_G$ just fixes the scale of the theory, and the other condition $h=0$ requires the colored
Higgs mass must be around the cutoff scale (the usual GUT scale). These are required even in the minimal SUSY
$SU(5)$ GUT, and therefore, in the natural GUT, the success of the gauge coupling unification in the minimal
SUSY $SU(5)$ GUT can be explained although the mass spectrum of superheavy particles does not respect
$SU(5)$. Although  $h$ is not zero and negative in the explicit model in Table \ref{SO10Higgs}, which is important to forbid the
Higgs mass term, the gauge coupling unification can be recovered by changing the $O(1)$ coefficients, for example, between 1/2 and 2.

In the following arguments, it is important that the cutoff scale $\Lambda$ must be around the usual GUT scale $\Lambda_G\sim 2\times 10^{16}$ GeV. 

\section{Spontaneous SUSY breaking in the natural GUT}
\label{sec:SSB}
In the natural GUT, VEVs of all negatively $U(1)_A$ charged fields except the SM doublet Higgs are determined by the $F$ flatness conditions for the positively $U(1)_A$ charged fields.  Therefore, if we decrease the number of negatively charged singlet fields or increase the number of positively charged singlet
fields, 
all the $F$ flatness conditions for the positively charged fields cannot be satisfied by fixing the VEVs of the negatively charged fields. 
Thus, we can realize an explicit GUT model with meta-stable vacuum which breaks SUSY 
spontaneously \cite{SUSYbreaking1, SUSYbreaking2}.
We note that some phenomenological requirements may prevent SUSY breaking spontaneously. 
In the subsection \ref{sec;breaking2}, we will discuss this problem in our model. 


\subsection{An explicit model}
One of the easiest way to build such a GUT model is to omit a negatively charged singlet $\bar Z$ from
the natural GUT in Table \ref{SO10Higgs}.  Then, one of the $F$ flatness conditions for $C'$ and $\bar C'$, i.e. $F_{C'}=0$ and $F_{\bar C'}=0$, cannot
be satisfied, and then SUSY is spontaneously broken. 
SUSY breaking scale can be obtained by
$m_{SUSY}\sim F_{\bar C'}/\Lambda\sim \lambda^{\bar c'+\frac{1}{2}(c-\bar c)}\Lambda\sim 4\times 10^{12}$
GeV, which is much higher than the electroweak scale. Of course, the SUSY breaking scale can be lower if we adopt
larger $\bar c'$ (or $c'$). For example, if we take $\bar c'=21(c'=18)$, $m_{SUSY}$ becomes about 2 TeV. 
Note that such a large $\bar c'$ allows the term $\bar C'AH^2C$, which results in the appearance of SUSY vacuum by fixing the VEV of $H^2$. 
We will discuss this issue at the end of this section.

Unfortunately, in this scenario, the gaugino masses are much smaller than $m_{SUSY}$.
Because of the large $U(1)_A$ charge of operator $\bar C'C$,
direct contribution from the term $\int d^2\theta \lambda^{\bar c'+c}\bar C'C W_A^\alpha W_{A\alpha}$ to the gaugino masses
becomes $m_{1/2}\sim m_{SUSY}^2/\Lambda$ which is much smaller than $m_{SUSY}$.
Contributions from gauge mediation and gaugino mediation give similar gaugino masses. 
This is because an approximate $U(1)_R$ symmetry in which positively (negatively) charged
fields have +2 (0) $U(1)_R$ charges appears in natural GUT at the meta-stable vacua, 
although this model originally has no $U(1)_R$ symmetry.
Indeed, the VEVs of negatively charged fields and $F$ of positively charged fields do not break the $U(1)_R$
symmetry. Therefore, to obtain the gaugino masses, the small $U(1)_R$ breaking like mass term of two
positively charged fields must be picked up in addition to the usual SUSY breaking factor.
As a result, the gaugino masses become very small in this scenario.

Let us estimate the contribution from the anomaly mediation \cite{anomalymed} to the gaugino masses.
Since the gravitino mass becomes $m_{3/2}\sim F_{\bar C'}/M_{Pl}\sim m_{SUSY}\Lambda/M_{Pl}$, where $M_{Pl}\sim 2\times 10^{18}$ GeV
 is the reduced Planck scale, the gaugino masses from the anomaly mediation are 
$
m_{1/2} \sim \frac{\alpha_ib_i}{4\pi}m_{3/2}\sim 10^{-4}m_{SUSY}.
$
If the gaugino masses are dominated by this contribution, very high-scale SUSY is required, which results in the fine-tuning.

Fortunately, according to the Ref. \cite{SUSYbreaking2}, in SUGRA, the constant term in the superpotential, which breaks $U(1)_R$
symmetry, can change the $U(1)_R$ symmetric vacuum and give the gaugino masses 
\begin{equation}
m_{1/2}\sim m_{3/2}\sim 10^{-2}m_{SUSY},
\end{equation} 
which improves the fine-tuning. 

For the sfermion masses, we have two main contributions. One of them gives $O(m_{SUSY})$ to the sfermion masses through the higher dimensional terms, for example, $\int d^4\theta |\bar C'|^2Q^\dagger Q$.
The other is called $D$-term contribution. 
In a simple model with one positively charged field 
in 
Refs. \cite{SUSYbreaking1, SUSYbreaking2}, the latter becomes about 10
times larger than the former (scenario A). However, in more realistic models with 
multiple positively charged fields, the hierarchy between the latter and the former becomes
milder, and the latter can be the same order of the former 
(scenario B) although some tuning between parameters may be required. 
These two scenarios
give different phenomenological consequences. 
Therefore, let us consider the phenomenological consequences in the following two
scenarios;
\begin{eqnarray}
{\rm A}&:& m_{1/2}\sim m_{3/2}\sim O(1\ {\rm TeV}), ~~ m_0^2\sim D_A\sim (10\ m_{SUSY})^2\sim O((1000\ {\rm TeV})^2), \\
{\rm B}&:& m_{1/2}\sim m_{3/2}\sim O(1\ {\rm TeV}), ~~ m_0^2\sim \ m_{SUSY}^2\sim O((100\ {\rm TeV})^2).
\end{eqnarray}
These can be realized if we take $\bar c'=18$.
Basically, SUSY contributions
to FCNC processes and CP violating processes are strongly suppressed because of large sfermion masses.

\subsection{Sfermion mass spectrum}
One of the most interesting features in scenario A is that the sfermion and Higgs mass squares are determined only by $D$-term contributions. 
Hereafter, we denote $D_A$ and $D_V$ are $D$-terms of $U(1)_A$ and $U(1)_V$ which is included in $SO(10)$ as $SO(10)\supset SU(5)\times U(1)_V$, respectively. 
Note that $\bf 16$ and $\bf 10$ of $SO(10)$ are divided under $SU(5)\times U(1)_V$ as
\begin{equation}
{\bf 16}={\bf 10}_{\frac{1}{5}}+{\bf\bar 5}_{-\frac{3}{5}}+1_1,\quad {\bf 10}={\bf 5}_{-\frac{2}{5}}+{\bf\bar 5}_{\frac{2}{5}},
\end{equation}
where the normalization of $U(1)_V$ is fixed as the $U(1)_V$ charge of the component with non-vanishing VEV is 1. Therefore, we obtain $i$-th generation sfermion masses $\tilde m_{10i}$ and $\tilde m_{{\bar 5}i}$ $(i=1,2,3)$, which are sfermion masses of $\bf 10$ fields and $\bf\bar 5$ fields, respectively. When $F$-term contributions for ${\bf 16}_i$ of $SO(10)$ are taken as $\tilde m_i$ and those for ${\bf 10}_T$ of $SO(10)$ are $m_T$, the sfermion masses
are given as follows:
\begin{eqnarray}
\tilde m_{10i}^2&=&\tilde m_i^2+\psi_ig_AD_A+\frac{1}{5}g_{10}D_V, \\
\tilde m_{{\bar 5}1}^2&=&\tilde m_1^2+\psi_1g_AD_A-\frac{3}{5}g_{10}D_V, \\ 
\tilde m_{{\bar 5}2}^2&=&\tilde m_T^2+tg_AD_A+\frac{2}{5}g_{10}D_V, \\ 
\tilde m_{{\bar 5}3}^2&=&\tilde m_2^2+\psi_2g_AD_A-\frac{3}{5}g_{10}D_V, 
\end{eqnarray}
where $g_A$ and $g_{10}$ are gauge couplings of $U(1)_A$ and $SO(10)$, respectively.
The soft SUSY breaking terms for the Higgs mass squared of up-type and down-type Higgs fields,
denoted as $m^2_{H_u}$ and $m^2_{H_d}$, are given by
\begin{eqnarray}
m^2_{H_u}&=&\tilde m_h^2+hg_AD_A-\frac{2}{5}g_{10}D_V, \\
m^2_{H_d}&=&\tilde m_h^2+hg_AD_A+\frac{2}{5}g_{10}D_V,
\end{eqnarray}
where $\tilde m_h^2$ is the $F$-term contribution to the Higgs field ${\bf 10}_H$ of 
$SO(10)$.
Note that the sfermion mass spectrum respects matter unification in $SU(5)$ GUT, i.e., sfermions which belong to an $SU(5)$
multiplet like $\bf 10$ of $SU(5)$ or $\bf\bar 5$ of $SU(5)$ have universal sfermion masses. On the other hand,
sfermions which belong to different multiplets of $SU(5)$ have generically different masses. Therefore, we can see the ``direct'' evidence of matter unification in $SU(5)$ GUT in the sfermion mass spectrum.  Although 100-1000 TeV is too 
high to be reached by near future experiments, it is much lower scale than the GUT scale. 
The usual renormalization group effects are very small except for $\tilde m_{{\bf 10}_3}$ fields in scenario B because the gaugino masses are much smaller than the sfermion 
masses and top Yukawa interaction has vanishing $U(1)_A$ charge to obtain large top Yukawa coupling.
Strictly speaking, since we have the stage with $SU(3)_C\times SU(2)_L\times SU(2)_R\times U(1)_{B-L}$,
the universal $D$-term contribution to sfermion masses in the same multiplet of $SU(5)$ splits because of the running gauge couplings as discussed in Ref. \cite{KMY}. 
We can obtain further information of the stage with
$SU(3)_C\times SU(2)_L\times SU(2)_R\times U(1)_{B-L}$ by measuring the splitting.
Interestingly, in the scenario A, the $D$-term contribution of $U(1)_V$ can be calculated as
\begin{equation}
g_{10}D_V=-\frac{\tilde m_{C}^2-\tilde m_{\bar C}^2}{2}=\frac{\bar c-c}{2}g_AD_A,
\end{equation}
because the SUSY breaking masses of the Higgs fields $C$ and $\bar C$ whose VEVs break 
$SU(2)_R\times U(1)_{B-L}$ into $U(1)_Y$ are determined by their $U(1)_A$ charges $c$ and $\bar c$ as
\begin{equation}
\tilde m_C^2= cg_AD_A, \quad \tilde m_{\bar C}^2=\bar cg_AD_A.
\end{equation}
Therefore, once we fix the natural GUT, the induced sfermion mass spectrum can be obtained in the scenario A. 
For example, in the explicit model in Table \ref{SO10Higgs}, the each predicted mass squared 
in the scenario A is obtained as
$(\tilde m_{{\bf 10}_1}^2, \tilde m_{{\bf 10}_2}^2,\tilde m_{{\bf 10}_3}^2, \tilde m_{{\bf\bar 5}_1}^2, \tilde m_{{\bf\bar 5}_2}^2, \tilde m_{{\bf\bar 5}_3}^2, m_{H_u}^2,m_{H_d}^2)=g_AD_A(48,38,18,36,31,26,-36,-24)/10$. On the other hand, in the scenario B, 
$g_{10}D_V$ is generically independent of $g_AD_A$ because of the $F$-term contribution
to these Higgs fields $C$ and $\bar C$.

When we take the $F$-term contributions to sfermion masses of ${\bf 27}_i$ as 
$\tilde m_i^2$ and those to Higgs field ${\bf 27}_H$ as $\tilde m_h^2$ in  the 
$E_6$ natural GUT\cite{BM}, the predicted sfermion and Higgs mass squares become
\begin{eqnarray}
\tilde m_{10i}^2&=&\tilde m_i^2+\psi_ig_AD_A+\frac{1}{5}g_{10}D_V+\frac{1}{4}g_6D_{V'}, \\
\tilde m_{{\bar 5}1}^2&=&\tilde m_1^2+\psi_1g_AD_A-\frac{3}{5}g_{10}D_V+\frac{1}{4}g_6D_{V'}, \\ 
\tilde m_{{\bar 5}2}^2&=&\tilde m_1^2+\psi_1g_AD_A+\frac{2}{5}g_{10}D_V-\frac{1}{2}g_6D_{V'}, \\ 
\tilde m_{{\bar 5}3}^2&=&\tilde m_2^2+\psi_2g_AD_A-\frac{3}{5}g_{10}D_V+\frac{1}{4}g_6D_{V'}, \\
m_{H_u}&=&\tilde m_h^2+hg_AD_A-\frac{2}{5}g_{10}D_V-\frac{1}{2}g_6D_{V'}, \\
m_{H_d}&=&\tilde m_h^2+hg_AD_A+\frac{2}{5}g_{10}D_V-\frac{1}{2}g_6D_{V'},
\end{eqnarray}
where $g_6$ and $D_{V'}$ are the gauge coupling constant of $E_6$ gauge group and the $D$-term of $U(1)_{V'}$,
which is included in $E_6$ as $E_6\supset SO(10)\times U(1)_{V'}$, respectively. Note that the fundamental
representation of $E_6$ is ${\bf 27}$ which is divided under $SO(10)\times U(1)_{V'}$ as
\begin{equation}
{\bf 27}={\bf 16}_{\frac{1}{4}}+{\bf 10}_{-\frac{1}{2}}+1_1,
\end{equation}
where the normalization of $U(1)_{V'}$ is fixed by the same way as $U(1)_V$. 
 In this $E_6$ natural GUT, 
the SM Higgs and GUT Higgs which breaks $E_6$ into $SO(10)$ are unified into ${\bf 27}_H$ (and ${\overline{\bf 27}}_{\bar H}$) of $E_6$. Again, the $D$-term contributions of $U(1)_{V'}$ and $U(1)_V$ in the scenario A can be calculated as
\begin{eqnarray}
g_{6}D_{V'}&=&-\frac{\tilde m_{H}^2-\tilde m_{\bar H}^2}{2}=\frac{\bar h-h}{2}g_AD_A, \\
g_{10}D_V&=&-\frac{\tilde m_{C}^2-\tilde m_{\bar C}^2}{2}=\frac{\bar c-c}{2}g_AD_A-\frac{1}{4}g_6D_{V'}
=\left(\frac{\bar c-c}{2}-\frac{\bar h-h}{8}\right)g_AD_A.
\end{eqnarray}
For example, if we take $(h,\bar h)=(-3,2)$ and $(c,\bar c)=(-4,0)$, we can obtain sfermion mass squares and Higgs mass square
as $(\tilde m_{{\bf 10}_1}^2, \tilde m_{{\bf 10}_2}^2,\tilde m_{{\bf 10}_3}^2, \tilde m_{{\bf\bar 5}_1}^2, \tilde m_{{\bf\bar 5}_2}^2, \tilde m_{{\bf\bar 5}_3}^2, m_{H_u}^2,m_{H_d}^2)=g_AD_A(54,44,24,43,38,33,-48,-37)/10$.
On the other hand, in scenario B, these $D$-terms are generically independent of each others and therefore the predictions for sfermion mass spectrum are not so
sharp as in scenario A.

Various natural GUT models with SUSY breaking scenarios can be tested by 
observing sfermion masses. 

\subsection{Long lived charged lepton}
Interestingly, this scenario predicts long lived charged lepton with odd 
R-parity. In the natural GUT, masses of all particles can be determined by their
$U(1)_A$ charges. 
When $\bar c'$ is taken to be large value as $\bar c'=18$, a pair of 
$E_R^c$ and $\bar E_R^c$ in $C'$ and $\bar C'$ becomes very light.
A mass term $\lambda^{c'+\bar c'}\bar C'C'$ gives them a mass 
$\lambda^{c'+\bar c'}\sim 200$ GeV when $(c',\bar c')=(3,18)$. 
Some enhancement factor for the mass is expected because many higher
dimensional terms like $\bar C'A^nZ^m(\bar CC)^lC'$ give the same 
contribution to the mass after developing the VEVs of $A$, $Z$, $C$, and
$\bar C$. Here we think the mass $O(1\ {\rm TeV})$. 


It is important to know what is the lightest MSSM SUSY particle (LMSP) to calculate the
lifetime of the long lived charged particle. 
In the scenario A, the Higgsino cannot be the LMSP 
because sufficiently large $\mu$ parameter is required to cancel the negative $D$-term
contributions to Higgs fields. So the bino is possible candidate for the LMSP.
However, in the scenario B, the Higgsino is possible to be the LMSP in addition to the bino.

\begin{figure}[htb]
  \begin{center}
    \includegraphics[width=0.45\textwidth,bb= 0 25 405 225]{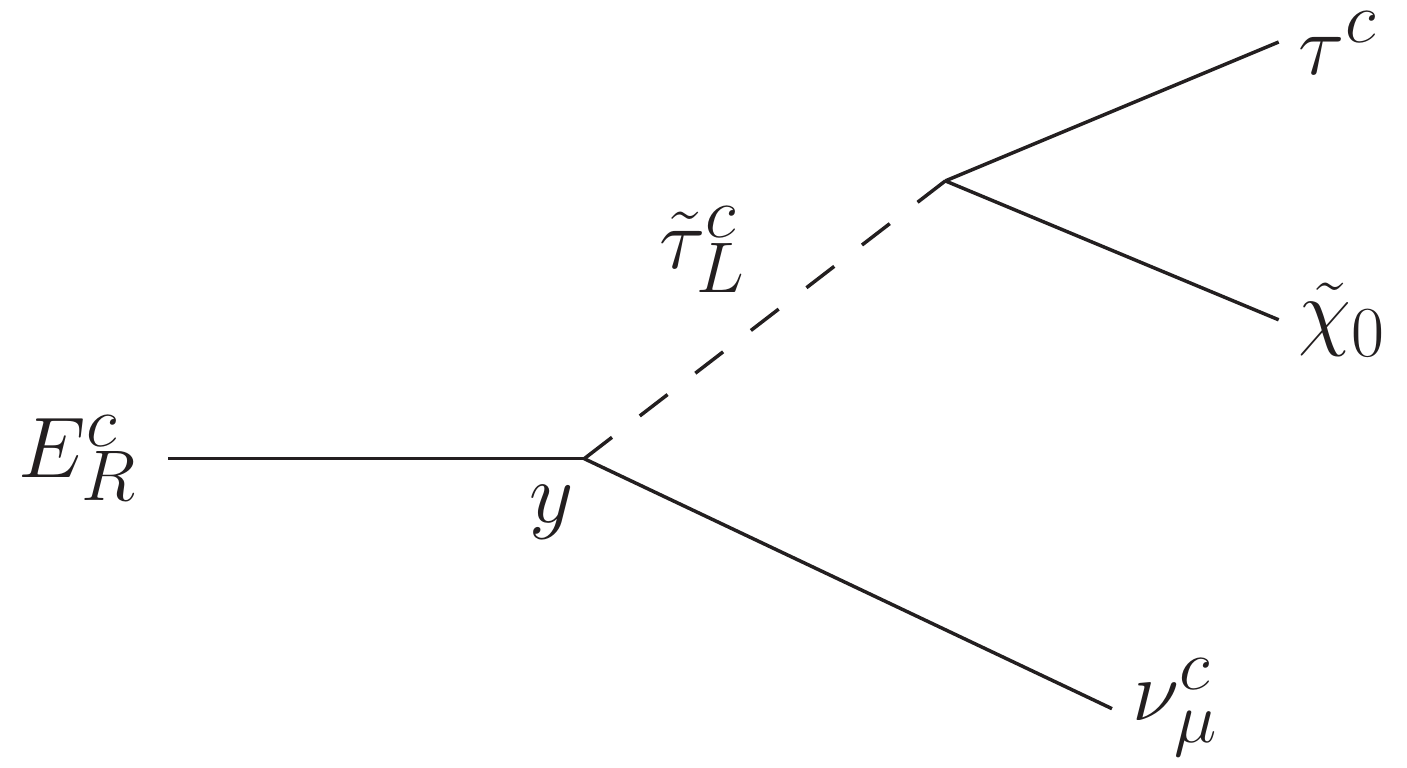}
    \caption{Feynman diagram which contributes the decay of long lived charged particle, $E_R^c$. $y$ denotes the Yukawa coupling of $C'\Psi_2T$ and its size can be estimated from their $U(1)_A$ charges as $y\sim \lambda^{c'+\psi_2+t}\sim 10^{-6}$.}
    \label{fig:ERdecay}
  \end{center}
\end{figure}

The lifetime of this long lived charged particle can be calculated through
the Feynman diagram in Fig.~\ref{fig:ERdecay} as
\begin{equation}
\tau_{E_R^c}\sim O(1) \, {\rm sec} \left(\frac{10^{-6}}{y}\right)^2\left(\frac{m_0}{1000\ {\rm TeV}}\right)^4\left(\frac{1\ {\rm TeV}}{m_{E_R^c}}\right)^5,
\end{equation}
if the bino is the lightest SUSY particle (LSP),
and 
\begin{equation}
\tau_{E_R^c}\sim O(0.1) \, {\rm sec} \left(\frac{10^{-6}}{y}\right)^2\left(\frac{\lambda^2}{y_\tau}\right)^2\left(\frac{m_0}{100\ {\rm TeV}}\right)^4\left(\frac{1\ {\rm TeV}}{m_{E_R^c}}\right)^5,
\end{equation}
if the Higgsino is the LSP.
Here $y_\tau$ is the Yukawa coupling of $\tau$ and
 $y$ is a Yukawa coupling of Yukawa interaction $C'\Psi_2T$ which is 
estimated as $y\sim \lambda^{c'+\psi_2+t}\sim 10^{-6}$. The success of the 
Big-Bang Nucleosynthesis (BBN) requires $\tau_{E_R^c}<1$ sec and LHC 
search gives the lower mass bound as
$m_{E_R^c}>574$ GeV \cite{LLCP}. LHC may find this long-lived charged lepton.


\subsection{Bino LMSP}
In the scenario A, the LMSP must be the bino as noted in the previous subsection.
If the LMSP becomes the lightest SUSY particle (LSP), 
the thermal relic abundance of the bino-like neutralino in the standard history of the universe
becomes much larger than the observed DM abundance because of quite
smaller annihilation cross section. 
Therefore, we need different thermal history than the standard one, for example, 
large entropy production etc.. 

However, we have the other possible candidates for the LSP in this scenario: LSP is gravitino or axino.
Since the gravitino mass is the same order of the gaugino masses, overproduction of the LMSP
cannot be improved if all the LMSP decays into the gravitino.
Besides, the lifetime of the LMSP becomes much longer than 1 second,
so that it spoils the success of the BBN. 
Therefore, we consider that the axino is the LSP. 

First, we explain where the axion is in our scenario.
 In the model with anomalous $U(1)_A$ gauge symmetry, the anomaly is cancelled by shift-transformation of a
moduli $M$ through the interactions 
\beq
{\cal L}_{\rm gauge}=\frac{1}{4\Lambda}\int d\theta^2\sum_ak_aM \, W_a^\alpha W_{a\alpha}+k_AM \,W_A^\alpha W_{A\alpha}+h.c.,
\eeq
where $W_a^\alpha$ and $W_A^\alpha$ are the superfield strength of 
$SO(10)$ and $U(1)_A$, and  $k_a$ and $k_A$ are
Kac-Moody levels of $SO(10)$ and $U(1)_A$, respectively. 
Here, $M$ is a moduli field. We note that $M$ is massless since there are no terms involving the moduli field in the superpotential.
On the other hand, the Nambu-Goldstone (NG) mode appears when 
negatively $U(1)_A$ charged fields develop non-vanishing VEVs because 
the superpotential is invariant under $U(1)_A$ symmetry. A linear 
combination of this NG mode and  the moduli is absorbed by the Higgs mechanism to make $U(1)_A$ 
gauge multiplet massive. The other 
combination becomes massless and can play the same role as an axion.
The fermionic partner of this axion is the axino.

It is important that the bino decays into photon and axino, and the lifetime of the bino can be shorter 
than 1 second as 
\begin{equation}
\tau_{\tilde{\chi}_0}\sim 0.1 \, {\rm sec} \left(\frac{1\ {\rm TeV}}{m_{\tilde{\chi}_0}}\right)^3
\left(\frac{\Lambda}{2\times 10^{16}\ {\rm GeV}}\right)^2.
\end{equation}
Therefore, the constraint from the BBN can be avoided.
Moreover, if the axino is sufficiently light,  the overclosure of the universe can be avoided.
Unfortunately, the axino produced by the decay of the bino cannot be cold DM because
the momentum is too large. Therefore, we need DM other than the LSP; for example, the axion. 
It is notable that the gravitino problem \cite{gravitino} can be solved without conflict with the 
BBN if the gravitino is lighter than the bino, because the gravitino decays into axino and axion \cite{AsakaYanagida}.

\subsection{The issues on the large $\bar c'$ case}
\label{sec;breaking2}
In our setup, 
the $F$ flatness conditions for $Z'$, $C'$ and $\bar C'$ only depend on $Z$, $A$, $C$ and $\bar C$.
$A$ is fixed by the $F$ flatness condition for $A'$ and the D flatness condition requires that the VEV of $C$ is equal to the one of $\bar C$. 
When our vacuum vanishes the $F$-term of $Z'$, either $F$-term of $C'$ or of $\bar C'$ is not vanishing and SUSY is broken, based on the argument in the subsection \ref{sec;breaking}.
This situation is, however, changed, if $\bar c'$ is large. For instance, 
additional $\bar C'H^2AC$ is allowed if $\bar c'=18$. Then,
the $F$ flatness condition for $\bar C'$ can be satisfied by obtaining non-vanishing VEV of $H^2$,
so that SUSY is not broken. 
To forbid this term, the maximal value of $\bar c'$ is 10, which results in too high SUSY breaking scale.
One possible way to introduce larger $\bar c'$ is  to introduce an additional discrete 
symmetry $Z_2'$ and two singlets. One singlet $S(s=-2,+)$ is odd $Z_2'$ parity and 
the other $S'(s'=4,+)$ is even whose $F$ flatness condition fixes the VEV of  $S$
as $\langle S\rangle\sim \lambda^{-s}$
via the superpotential $W_{S'}=\lambda^{s'}S'+\lambda^{s'+2s}S'S^2$.
When  $\bar C'$ has odd $Z_2'$ parity, then the term $\bar C'H^2AC$ is forbidden
even if $\bar c'=12$. Unfortunately, if we take $\bar c'$ larger than 12, the term
$\bar C'H^2ASC$ is allowed, which develops non-vanishing VEV of $H^2$. Note that
it is not workable to adopt $s\leq-3$ because the term $S'H^2$ must be forbidden.
Therefore, $\bar c'=18$ becomes possible when we introduce four additional $Z_2^i$
$(i=1,2,3,4)$ 
symmetries and eight singlets: four $S^i(s^i=-2,+)$ and four $S'^i(s'^i=4,+)$. Here, $S'^i$ has even parity
for all $Z_2$ symmetries, while $S^i$ is odd for $Z_2^i$ but even for $Z_2^j$ $(j\neq i)$.
$\bar C'$ has odd parity under all $Z_2^i$ symmetries. 

We have some difficulty in solving the $\mu$ problem in the scenario A. 
We need sufficiently large $\mu$
 parameter so that large negative $D$-term contribution to Higgs mass squares can be
 cancelled
by the $\mu$. However, in the solution discussed in Ref. \cite{mu}, $\mu$ is proportional
to $A$-term which is of order the gravitino mass $m_{3/2}\sim O(1)$ TeV. Moreover, 
we need the superpotential as
\begin{equation}
W_{\tilde S'}=\lambda^{\tilde s'}\tilde S'+\lambda^{\tilde s'+\tilde s}(\tilde S'\tilde S)+\lambda^{\tilde s'+2h}(\tilde S'H^2)
\end{equation}
to solve the $\mu$ problem. $F$-flatness condition of positively charged field $\tilde S'$
 determines the VEV of the negatively charged field
$\tilde S$. Obviously, the filed
$\tilde S$ must not be included in the $F$-flatness condition of $\bar C'$ to break SUSY 
spontaneously. If $\tilde s=-6$ or smaller, the above requirement is satisfied. However,
$\tilde s=-6$ results in too small $\mu$ because  $\mu$ is given by
\begin{equation}
\mu=\lambda^{2h-2\tilde s}m_{3/2}.
\end{equation}
 One way to avoid this situation is to introduce another discrete symmetry, for example,
$Z_6$, and $\tilde S$ is 1 under the $Z_6$. Then the superpotential becomes
\begin{equation}
W_{\tilde S'}=\lambda^{\tilde s'}\tilde S'+\lambda^{\tilde s'+6\tilde s}(\tilde S'\tilde S^6)+\lambda^{\tilde s'+2h}(\tilde S'H^2),
\end{equation}
and $\mu$ can be large as $\lambda^{-4}m_{3/2}\sim 500$ TeV if we take $\tilde s=-1$.
Unfortunately, this solution leads to too large $B\mu$ term as 
\begin{equation}
B\mu\sim \lambda^{2h-2\tilde s}m_{\tilde S}^2,
\end{equation}
where  the soft SUSY breaking mass square of the singlet $\tilde S$ is proportional to the 
$D$-term.  Introducing two pairs of $\tilde S'$ and $\tilde S$, we can avoid this problem, although some tuning between parameters is required. 
.

On the other hand, the scenario B has no such difficulties although some tuning is needed
to obtain $D_A\sim |F_{\bar C'}|^2$.


\section{summary and discussion}
\label{sec:summary}
We have proposed a simple GUT scenario in which SUSY is spontaneously broken in the visible sector. 
We have started with the natural GUT, in which almost all problems including the doublet-triplet splitting problem can be solved under the natural assumption that all interactions allowed by
the symmetry are introduced with O(1) coefficients.
Interestingly, only small deviation from the natural GUT can realize spontaneous SUSY breaking.
Concretely, spontaneous SUSY breaking have been realized by omitting one singlet field from the
natural GUT model with SUSY vacuum. 

This scenario predicts high-scale SUSY in which sfermion masses and Higgs masses are O(100)-O(1000) TeV, while gaugino masses are O(1) TeV. This high-scale SUSY solves the SUSY flavor problem and the SUSY CP problem, although the fine-tuning in Higgs sector is not avoidable. 
We have discussed two scenarios for SUSY breaking. In the scenario A, $D$-term 
contributions to sfermion masses dominate, and the sfermion mass scale is around
1000 TeV. In the scenario B, the $F$-term contributions and the $D$-term contributions
are comparable. 
We have found interesting predictions in these scenarios.
Since the sfermion mass spectrum is respected to $SU(5)$ matter unification, 
the signature of the GUT appears in sfermion mass spectrum. Furthermore, superheavy long-lived charged lepton is predicted, which may be seen in the LHC. 

In the scenario A, the LMSP must be the bino, whose thermal production density is much larger than
the observed DM density. The axino LSP can solve this issue as well as the gravitino problem.

\section*{acknowledgments}
N.M. thanks K.~Harigaya for valuable communication on DM. He also thanks N.~Nagata for information
on long-lived charged particle.  This work is supported in part by the Grant-in-Aid for Scientific Research 
  from the Ministry of Education, Culture, Sports, 
 Science and Technology in Japan  No.~19K03823(N.M.), No. 19H04614, No. 19H05101 and No. 19K03867(Y.O.).


\begin{thebibliography}{99}
\bibitem{GG}
  H.~Georgi and S.~L.~Glashow,
  Phys.\ Rev.\ Lett.\  {\bf 32}, 438 (1974).
\bibitem{DTS}
For the review,
  L.~Randall and C.~Csaki,
  In *Palaiseau 1995, SUSY 95* 99-109
  [hep-ph/9508208].

\bibitem{naturalGUT}
  N.~Maekawa,
  Prog.\ Theor.\ Phys.\  {\bf 106}, 401 (2001)
  [hep-ph/0104200];
  N.~Maekawa and T.~Yamashita,
  Prog.\ Theor.\ Phys.\  {\bf 107}, 1201 (2002).
  [hep-ph/0202050].

\bibitem{anomalousU(1)}
  E.~Witten,
  Phys.\ Lett.\  {\bf 149B}, 351 (1984);
  M.~Dine, N.~Seiberg and E.~Witten,
  Nucl.\ Phys.\ B {\bf 289}, 589 (1987);
  J.~J.~Atick, L.~J.~Dixon and A.~Sen,
  Nucl.\ Phys.\ B {\bf 292}, 109 (1987);
  M.~Dine, I.~Ichinose and N.~Seiberg,
  Nucl.\ Phys.\ B {\bf 293}, 253 (1987).

\bibitem{FI}
  P.~Fayet and J.~Iliopoulos,
  Phys.\ Lett.\  {\bf 51B}, 461 (1974);
  P.~Fayet,
  Nucl.\ Phys.\ B {\bf 90}, 104 (1975).

\bibitem{GS} 
  M.~B.~Green and J.~H.~Schwarz,
  Phys.\ Lett.\  {\bf 149B}, 117 (1984).

\bibitem{DW}
  S.~Dimopoulos and F.~Wilczek, NSF-ITP-82-07;
  M.~Srednicki,
  Nucl.\ Phys.\ B {\bf 202}, 327 (1982).

\bibitem{Barr:1997hq} 
  S.~M.~Barr and S.~Raby,
  Phys.\ Rev.\ Lett.\  {\bf 79}, 4748 (1997)
  [hep-ph/9705366].

\bibitem{gcu} 
  N.~Maekawa,
  Prog.\ Theor.\ Phys.\  {\bf 107}, 597 (2002)
  [hep-ph/0111205];
  N.~Maekawa and T.~Yamashita,
  Phys.\ Rev.\ Lett.\  {\bf 90}, 121801 (2003)
  [hep-ph/0209217].

\bibitem{SUSYbreaking1}
  S.-G.~Kim, N.~Maekawa, H.~Nishino and K.~Sakurai,
  Phys.\ Rev.\ D {\bf 79}, 055009 (2009)
  [arXiv:0810.4439 [hep-ph]].

\bibitem{SUSYbreaking2}
  N.~Maekawa, Y.~Omura, Y.~Shigekami and M.~Yoshida,
  Phys.\ Rev.\ D {\bf 97}, no. 5, 055015 (2018)
  [arXiv:1712.05107 [hep-ph]].

\bibitem{anomalymed}
  L.~Randall and R.~Sundrum,
  Nucl.\ Phys.\ B {\bf 557}, 79 (1999)
  [hep-th/9810155]; 
  G.~F.~Giudice, M.~A.~Luty, H.~Murayama and R.~Rattazzi,
  JHEP {\bf 9812}, 027 (1998)
  [hep-ph/9810442].

\bibitem{KMY}
  Y.~Kawamura, H.~Murayama and M.~Yamaguchi,
  Phys.\ Rev.\ D {\bf 51}, 1337 (1995)
  [hep-ph/9406245].

\bibitem{BM}
  M.~Bando and N.~Maekawa,
  Prog.\ Theor.\ Phys.\  {\bf 106}, 1255 (2001)
  [hep-ph/0109018].

\bibitem{LLCP}
  S.~Chatrchyan {\it et al.} [CMS Collaboration],
  JHEP {\bf 1307}, 122 (2013)
  [arXiv:1305.0491 [hep-ex]];
  V.~Khachatryan {\it et al.} [CMS Collaboration],
  Phys.\ Rev.\ D {\bf 94}, no. 11, 112004 (2016)
  [arXiv:1609.08382 [hep-ex]];
  M.~Aaboud {\it et al.} [ATLAS Collaboration],
  Phys.\ Rev.\ D {\bf 99}, no. 9, 092007 (2019)
  [arXiv:1902.01636 [hep-ex]].


\bibitem{mu}
  N.~Maekawa,
  Phys.\ Lett.\ B {\bf 521}, 42 (2001)
  [hep-ph/0107313].
\bibitem{gravitino}
  S.~Weinberg,
  Phys.\ Rev.\ Lett.\  {\bf 48}, 1303 (1982);
  M.~Y.~Khlopov and A.~D.~Linde,
  Phys.\ Lett.\ B {\bf 138}, 265 (1984);
  J.~R.~Ellis, J.~E.~Kim and D.~V.~Nanopoulos,
  Phys.\ Lett.\ B {\bf 145}, 181 (1984);
%
%
  M.~H.~Reno and D.~Seckel,
  Phys.\ Rev.\ D {\bf 37}, 3441 (1988);
%
%
  M.~Kawasaki and T.~Moroi,
  Prog.\ Theor.\ Phys.\  {\bf 93}, 879 (1995)
  [hep-ph/9403364, hep-ph/9403061];
%
%
  K.~Kohri,
  Phys.\ Rev.\ D {\bf 64}, 043515 (2001)
  [astro-ph/0103411];
%
%
  M.~Kawasaki, K.~Kohri and T.~Moroi,
  Phys.\ Rev.\ D {\bf 71}, 083502 (2005)
  [astro-ph/0408426];
%
%
  M.~Kawasaki, K.~Kohri, T.~Moroi and A.~Yotsuyanagi,
  Phys.\ Rev.\ D {\bf 78}, 065011 (2008)
  [arXiv:0804.3745 [hep-ph]].

\bibitem{AsakaYanagida}
  T.~Asaka and T.~Yanagida,
  Phys.\ Lett.\ B {\bf 494}, 297 (2000)
  [hep-ph/0006211].
\end{thebibliography}
\end{document}